\documentstyle[12pt]{article} 
\advance \topmargin by -\headsep
\evensidemargin \oddsidemargin
\title{The Principle of Non-Gravitating Vacuum Energy and some of its 
consequences}
\author{E.I.Guendelman \thanks{GUENDEL@BGUmail.BGU.AC.IL} and
        A.B.Kaganovich \thanks{ALEXK@BGUmail.BGU.AC.IL}}
\date {Physics Department, Ben Gurion University of the Negev, 
   Beer  Sheva 84105, Israel} 

\begin{document}
\maketitle
\begin{abstract}
For Einstein's General Relativity (GR) or the alternatives suggested up to date, 
the 
vacuum energy gravitates.  We present a model where a new measure is 
introduced for 
integration of the total action in the D-dimensional space-time. This measure 
is built from D scalar fields $\varphi_{a}$. As a consequence of such a choice
 of the measure, the matter lagrangian $L_{m}$ can be changed by adding 
 a constant while no gravitational effects, 
like a cosmological term, are induced. Such Non-Gravitating Vacuum Energy 
(NGVE) theory has an 
infinite dimensional symmetry group which contains volume-preserving 
diffeomorphisms in the internal space of scalar fields  $\varphi_{a}$. 
Other symmetries contained in this symmetry group, suggest a deep 
connection of this theory with theories of extended objects.  In general 
{\em the theory is different from GR} although for certain choices of $L_{m}$,
 which are related to the 
existence of an additional symmetry, solutions of GR are 
solutions of the model.  This is achieved in four dimensions if $L_{m}$ is 
due 
to fundamental bosonic and fermionic strings.  Other types of 
matter where 
this feature of the theory is realized, are for example: scalars without
 potential or 
subjected to nonlinear constraints, massless fermions and point particles. 
The point particle plays a special role, since it is a
good phenomenological description of matter at large distances. de Sitter 
space is realized in an unconventional way, where the de Sitter metric 
holds, but such de Sitter space is supported by the existence of a 
variable scalar field  which in practice destroys the maximal symmetry.  
The 
only space - time where maximal symmetry is not broken, in a dynamical 
sense, is 
Minkowski space. The theory has non trivial dynamics in 1+1 dimensions, 
unlike GR. 
\end{abstract}

\pagebreak

\section{Introduction}

\bigskip

As it is known, in the general theory of relativity 
energy is a source of gravity, which is described by the 
metric tensor $g_{\mu\nu}$. This makes an important difference to ideas 
developed for flat space physics where the origin with respect to which 
we measure energy does not matter, that is the energy is defined up to an 
additive constant. For general relativity in contrast, all the energy has 
a gravitational effect, therefore the origin with respect to which we 
define the energy is important.

In quantum mechanics, there is the so called zero-point energy associated 
to the zero-point fluctuations. In the case of quantum fields, such 
zero-point fluctuations turn out to have an associated energy density 
which is infinite. In fact there is a zero point vacuum energy - momentum tensor 
of the form $T_{\mu\nu}^{vac}=A\eta_{\mu\nu}$ in flat space 
($\eta_{\mu\nu}$ is the Minkowski metric), or 
$T_{\mu\nu}^{vac}=Ag_{\mu\nu}+(terms \propto R)$ in curved space. Here $A$ 
is infinite.

Notice that the appearance of an energy-momentum tensor proportional to 
$g_{\mu\nu}$ in Einstein's equations is equivalent\cite{Zel} to what 
Einstein called the `'cosmological constant`` or `'$\Lambda$- 
term``. It was introduced by Einstein\cite{Ein} in the form
\begin{equation}
 R_{\mu\nu}-\frac{1}{2}g_{\mu\nu}R-\Lambda g_{\mu\nu}=
 \frac{\kappa}{2}T_{\mu\nu}  	
   	\label{Ein}
\end{equation}

Such $\Lambda$-term does not violate any known symmetry. Therefore, 
normally we would not consider excluding it, if we were to apply the 
arguments usually made in quantum field theory. However, we get into 
trouble if we note that the natural scale of such a term, obtained on
dimensional grounds, is of the order of magnitude of the Planck density.
The problem is more severe once we realize that the zero point vacuum energy 
appears as an infinite quantity.

Indeed, in order to get agreement with observations, different 
sources of energy density have to compensate with each other almost 
exactly to high 
accuracy, thus creating an acute 
`'fine tuning problem``.

In order to explain this so called `'cosmological constant problem``, a 
variety of ideas have been developed, see for example 
reviews\cite{reviews}. Among these attempts, possible changes in  
gravity theory were studied\cite{reviews}, where the result was that 
the cosmological constant appeared as an integration constant, for 
example, in `'nondynamical $\sqrt{-g}$`` models. The reason why such a 
constant should be picked zero is unclear however.

In this paper we will also suggest a modification of gravity theory by 
imposing the principle that the vacuum energy density, to be identified 
with the constant part of the lagrangian density, should not contribute 
to the equation of motion. The realization of this idea in the model 
considered here, apart from leading to a geometrically interesting new 
theory, also leads to the possibility of new gravitational effects.

\bigskip

\section{The Model}

\bigskip
\subsection{A new measure and generally coordinate invariant action}

All approaches to the cosmological constant problem have been made 
under  the assumption that 
the invariant measure to be used for integrating the total lagrangian 
density in the action is $\sqrt{-g}d^Dx$.  In the present paper, this 
particular assumption will be modified, and the result will be that, by an 
appropriate generally coordinate invariant choice of the measure, the 
theory will not be sensitive to a change in the lagrangian density by the 
addition of a constant, in contrast to the Einstein-Hilbert action, where 
such a change generates a cosmological term.
    
Let the measure of integration in a D dimensional space-time be chosen as 
 $\Phi d^Dx$, where $\Phi$ is a yet unspecified scalar density of weight 1. 
In order  to 
achieve the result that the vacuum energy does not gravitate, we will start 
from the demand that the addition of a constant in the lagrangian density 
does not affect the dynamics of the theory. This means that $\int L\Phi d^Dx$ 
and $\int (L+constant)\Phi d^Dx$ must reproduce the same equations of motion. 
This is of course achieved if $\Phi$ is a total derivative.

The simplest choice for a scalar density of weight 1, which is as well a total 
derivative, may be realized by using D scalar fields $\varphi_{a}(x)$ and 
then defining 
\begin{equation}
\Phi \equiv \varepsilon_{a_{1}a_{2}\ldots a_{D}}
\varepsilon^{\alpha_{1}\alpha_{2}\ldots \alpha_{D}} 
(\partial_{\alpha_{1}}\varphi_{a_{1}}) 
(\partial_{\alpha_{2}}\varphi_{a_{2}}) \ldots 
(\partial_{\alpha_{D}}\varphi_{a_{D}})	\label{Fish}
\end{equation} 

Here $\varepsilon^{\alpha_{1}\alpha_{2}\ldots \alpha_{D}}=1$ if 
$\alpha_{1}=0,\alpha_{2}=1\ldots \alpha_{D}=D-1$ and $\pm1$ 
according to whether $(\alpha_{1},\alpha_{2},\ldots,\alpha_{D})$ is an even 
or an odd permutation of $(0,1,\ldots,D-1)$.  Likewise for 
$\varepsilon_{a_{1}a_{2}\ldots a_{D}}$  $(a_{i}= 1,2,\ldots, D)$  
Therefore the total action we will consider is 
\begin{equation}
S = \int(- \frac{1}{\kappa}R + L_{m})\Phi d^{D}x
	\label{Action}
\end{equation}
where $\kappa=16\pi G$, $R$ is the scalar curvature and we will take $L_{m}$ 
not to depend on any of the scalars $\varphi_{a}(x)$.  Notice that if we 
consider parity or time reversal transformations, then $S \rightarrow 
-S$, which does not affect the classical equations of motion.  
Quantum mechanically (considering for example the path integral 
approach), such transformation will transform {\em total} amplitudes into
their complex conjugates, therefore leaving probabilities unchanged.
 
Notice that $\Phi$ is the jacobian of the mapping 
$\varphi_{a}=\varphi_{a}(x^{\alpha})$, $a=1,2,\ldots,D$. If 
this mapping is non singular $(\Phi\neq 0)$ then (at least locally) there 
is  the inverse mapping $x^{\alpha}=x^{\alpha}(\varphi_{a})$, $\alpha= 
0,1\ldots,D-1$.  Since $\Phi d^{D}x = D! 
d\varphi_{1}\wedge d\varphi_{2}\wedge\ldots\wedge d\varphi_{D}$  we can 
think  $\Phi d^{D}x$ as integrating in the internal space 
variables $\varphi_{a}$.  Besides, if $\Phi \neq 0$ then there is a 
coordinate frame where the coordinates are the scalar fields themselves.

The field $\Phi$ is invariant under the volume preserving diffeomorphisms 
in internal space: $\varphi^{\prime}_{a}=\varphi^{\prime}_{a}(\varphi_{b})$ 
where \begin{equation}
\varepsilon_{a_{1}a_{2}\ldots 
a_{D}}\frac{\partial{\varphi^{\prime}}_{b_{1}}}{\partial{\varphi}_{a_{1}}}
\frac{\partial{\varphi^{\prime}}_{b_{2}}}{\partial{\varphi}_{a_{2}}}\ldots
\frac{\partial{\varphi^{\prime}}_{b_{D}}}{\partial{\varphi}_{a_{D}}}=
 \varepsilon_{b_{1}b_{2}\ldots b_{D}}
 	\label{epsilon}
\end{equation}
 
 \subsection{Equations of motion}
 
The equations of motion obtained by variation of the action (\ref{Action}) 
with respect to the scalar fields $\varphi_{b}$ are
\begin{equation}
 A_{b}^{\mu}\partial_{\mu}(-\frac{1}{\kappa}R + L_{m} ) = 0 
 \label{Inv}
 \end{equation}
  where
  \begin{equation}  	
  A_{b}^{\mu}\equiv \varepsilon_{a_{1}a_{2}\ldots a_{D-1}b}
\varepsilon^{\alpha_{1}\alpha_{2}\ldots \alpha_{D-1}\mu} 
(\partial_{\alpha_{1}}\varphi_{a_{1}}) 
(\partial_{\alpha_{2}}\varphi_{a_{2}}) \ldots 
(\partial_{\alpha_{D-1}}\varphi_{a_{D-1}})
	\label{A}
  \end{equation}   
It follows from (\ref{Fish}) that  
$A_{b}^{\mu}\partial_{\mu}\varphi_{b^{\prime}}=D^{-1}\delta_{bb^{\prime}}\Phi$ 
and taking the determinant of 
both sides, we get $det (A_{b}^{\mu}) = \frac{D^{-D}}{D!}\Phi^{D-1}$.  
Therefore if $\Phi\neq 0$, which we will assume in what follows, the 
only solution for (\ref{Inv}) is
\begin{equation} 
- \frac{1}{\kappa} R + L_{m} = constant\equiv M \label{IntM}
\end{equation}

Variation of $S_{g}\equiv-\frac{1}{\kappa}\int R\Phi d^{D}x$ 
with respect to 
$g^{\mu\nu}$ leads to the result\cite{Chern}
\begin{equation}
	\delta S_{g}=-\frac{1}{\kappa}\int \Phi[R_{\mu\nu}+
	(g_{\mu\nu}\Box-\nabla_{\mu}\nabla_{\nu})]\delta g^{\mu\nu}d^{D}x
	\label{Andro}
\end{equation}
In order to perform the correct integration by parts we have to make 
use 
the scalar field $\chi\equiv\frac{\Phi}{\sqrt{-g}}$, which is invariant 
under continuous general coordinate transformations, instead of the 
scalar density 
$\Phi$.  Then integrating by parts and ignoring a total derivative term 
which has the form $\partial_{\alpha}(\sqrt{-g}P^{\alpha})$, where 
$P^{\alpha}$ is a vector field, we get
\begin{equation}
	\frac{\delta S_{g}}{\delta g^{\mu\nu}}=-\frac{1}{\kappa}\sqrt{-g}[\chi 
	R_{\mu\nu}+g_{\mu\nu}\Box\chi-\chi_{,\mu;\nu}]
	\label{var}
\end{equation}
In a similar way varying the matter part of the action (\ref{Action}) 
with 
respect to $g^{\mu\nu}$ and making use the scalar field $\chi$ we can 
express a result in terms of the standard matter energy-momentum tensor 
$T_{\mu\nu}\equiv
\frac{2}{\sqrt{-g}}\frac{\partial(\sqrt{-g}L_{m})}{\partial 
g^{\mu\nu}}$. Then after some algebraic manipulations we get instead of 
Einstein's equations
\begin{equation}
	G_{\mu\nu}=\frac{\kappa}{2} [T_{\mu\nu} - 
	\frac{1}{2}g_{\mu\nu}(T^{\alpha}_{\alpha}+(D-2)L_{m})]+
	\frac{1}{\chi}\left(\frac{D-3}{2}g_{\mu\nu}\Box\chi+\chi_{,\mu;\nu}\right)
	    	\label{Ein}
\end{equation}
where $G_{\mu\nu}\equiv R_{\mu\nu}-\frac{1}{2}Rg_{\mu\nu}$. 
 
By contracting (\ref{Ein}) and using (\ref{IntM}), we get
\begin{equation}
\Box\chi-\frac{\kappa}{D-1}[M+\frac{1}{2}(T^{\alpha}_{\alpha} 
+(D-2)L_{m})]\chi=0		
\label{Scal}
\end{equation}

By using eq.(\ref{Scal}) we can now exclude  
$T^{\alpha}_{\alpha}+(D-2)L_{m}$ from eq.(\ref{Ein}):
\begin{equation}
G_{\mu\nu}=\frac{\kappa}{2} [T_{\mu\nu}+Mg_{\mu\nu}]+
	\frac{1}{\chi}[\chi_{,\mu;\nu}-g_{\mu\nu}\Box\chi]
	\label{M-eq}
\end{equation}
Notice that eqs.(\ref{Inv}) and (\ref{Ein}) are invariant under the 
addition 
to $L_{m}$ a constant piece, since the combination $T_{\mu\nu}-
\frac{1}{2}g_{\mu\nu}[T^{\alpha}_{\alpha}+(D-2)L_{m}]$ is invariant.  
However 
by fixing the constant part of the difference $L_{m}-M$ in the solution 
(\ref{IntM}) of eq.(\ref{Inv}), we break this invariance. {\em To give a 
definite
physical meaning to the integration constant $M$, we conventionally 
take the constant part of $L_{m}$ equal to zero \/}.

It is very important to note that the terms depending on the matter 
fields 
in eq.(\ref{M-eq}) as well as in eq.(\ref{Ein}) do not contain $\chi$ 
-field, in contrast to the usual scalar-tensor theories, like 
Brans-Dicke 
theory.  As a result of this feature of the NGVE- theory, 
the 
gravitational constant does not suffer space-time variations.  
However, the matter 
energy-momentum tensor $T_{\mu\nu}$ is not conserved.  Actually, taking the 
covariant 
divergence of both sides of eq.(\ref{M-eq}) and using the identity 
$\chi^{\alpha} _{;\nu;\alpha}=(\Box
\chi),_{\nu}+\chi^{\alpha}R_{\alpha\nu}$, eqs.(\ref{M-eq}) and 
(\ref{Scal}), 
we get the equation of matter non conservation
\begin{equation}
T_{\mu\nu}\:^{;\mu}=-2\frac{\partial L_{m}}{\partial g^{\mu\nu}}
g^{\mu\alpha}\partial_{\alpha}ln\chi 
\label{Noncons}
\end{equation}

\subsection{Volume preserving symmetries and associated conserved 
quantities}

From the volume preserving 
symmetries $\varphi^{\prime}_{a}=\varphi^{\prime}_{a}(\varphi_{b})$ 
defined by eq.(\ref{epsilon}) which for the infinitesimal case implies
\begin{equation}
	\varphi^{\prime}_{a}=\varphi_{a}+
	\lambda\varepsilon^{aa_{1}\ldots a_{D}}
	\frac{\partial F_{a_{1}a_{2}\ldots a_{D-1}}(\varphi_{b})}
	{\partial\varphi_{a_{D}}}
	\label{infphi}
\end{equation}
($\lambda\ll 1$), we obtain, through Noether's theorem the following 
conserved quantities
\begin{equation}
	j_{V}^{\mu}=A_{a}^{\mu}(-\frac{1}{\kappa}R+L_{m})
	\varepsilon_{aa_{1}\ldots a_{D}}
	\frac{\partial F_{a_{1}a_{2}\ldots a_{D-1}}(\varphi_{b})}
	{\partial\varphi_{a_{D}}}
	\label{jv}
\end{equation}

\subsection{Symmetry transformations with the total lagrangian density as 
a parameter and associated conserved quantities}

Let us consider the following infinitesimal shift of the fields 
$\varphi_{a}$ by an arbitrary infinitesimal function of the total lagrangian
density 
$L\equiv-\frac{1}{\kappa}R+L_{m}$, that is
\begin{equation}
   	\varphi^{\prime}_{a}=\varphi_{a}+\epsilon g_{a}(L),   \epsilon\ll 1
   	\label{L}
   	\end{equation}
In this case the action is transformed according to
\begin{equation}
	\delta S=\epsilon D\int A_{a}^{\mu}L\partial_{\mu} 
	g_{a}(L)d^{D}x=\epsilon\int\partial_{\mu}\Omega^{\mu}d^{D}x
	\label{deltaS}
\end{equation}
where $\Omega^{\mu}\equiv DA_{a}^{\mu}f_{a}(L)$ and $f_{a}(L)$ being defined 
from $g_{a}(L)$ through the equation $L\frac{dg_{a}}{dL}=\frac{df_{a}}{dL}$. 
To obtain the last expression in the equation(\ref{deltaS}) it is 
necessary to note that $\partial_{\mu}A_{a}^{\mu}\equiv 0$. By means of 
the Noether's theorem, this symmetry leads to the conserved current
\begin{equation}
	j_{L}^{\mu}=A_{a}^{\mu}(Lg_{a}-f_{a})\equiv
	A_{a}^{\mu}\int_{L_{0}}^{L}g_{a}(L^{\prime})dL^{\prime}
	\label{jl}
\end{equation}

For certain matter models, the symmetry structure is even richer (see 
discussion of the `'Einstein symmetry `` in the next 
section).  The complete understanding of the group structure and the consequences of 
these symmetries is not known to us at present, but we will expect to 
report on this in future publications.  	
   	
\bigskip

\section{Einstein symmetry and Einstein sector of solutions}

\subsection{Einstein condition}

We are interested now in studying the question whether there is an 
Einstein 
sector of solutions, that is are there solutions that satisfy 
Einstein's 
equations? First of all we see that eqs.(\ref{M-eq}) coincide with 
Einstein's equations with cosmological constant ${\kappa}M$ only if the
$\chi$-field is a constant.  From eq.(\ref{Scal}) we conclude that this is 
possible 
only if an essential restriction on the matter model is imposed: $2M + 
T^{\alpha}_{\alpha}+(D-2)L_{m}\equiv 2[M+g^{\mu\nu}\frac{\partial 
L_{m}}{\partial g^{\mu\nu}}-L_{m}]=0$.  However we have not found 
any 
reasonable matter model where this condition is satisfied for $M\neq 0$ (recall 
that 
the constant part of $L_{m}$ is taken to be zero).  In the case $M=0$ 
this 
condition becomes
\begin{equation}
	g^{\mu\nu}\frac{\partial L_{m}}{\partial g^{\mu\nu}}-L_{m}=0
	\label{Sector}
\end{equation}
which means that $L_{m}$ is an homogeneous function of $g^{\mu\nu}$ of 
degree one, in any dimension.  If condition (\ref{Sector}) is satisfied 
then the equations of motion allow solutions of GR to be solutions of 
the 
model, that is $\chi = constant$ and $G_{\mu\nu}=
\frac{\kappa}{2}T_{\mu\nu}$.

\subsection{Einstein symmetry}

It is interesting to observe that when condition (\ref{Sector}) is 
satisfied, a new symmetry of the action(\ref{Action}) appears. We will call 
this symmetry `'Einstein 
symmetry`` (because (\ref{Sector}) leads to the existence of an Einstein 
sector of solutions).
Such symmetry consists of the scalings
\begin{equation}
	g^{\mu\nu}\rightarrow \lambda g^{\mu\nu}
	\label{Eg}
\end{equation}
\begin{equation}
	\varphi_{a}\rightarrow \lambda^{-\frac{1}{D}}\varphi_{a}
	\label{Ephi}
\end{equation}
To see that this is indeed a symmetry, note that from definition of scalar 
curvature it follows that $R\rightarrow \lambda 
R$ when the transformation (\ref{Eg}) is performed. Since condition 
(\ref{Sector}) means that $L_{m}$ is a homogeneous function of 
$g^{\mu\nu}$ of degree 1, we see that under the transformation 
(\ref{Eg}) the matter lagrangian $L_{m}\rightarrow \lambda L_{m}$. From 
this we conclude that (\ref{Eg})-(\ref{Ephi}) is indeed a symmetry of the 
action(\ref{Action}) when (\ref{Sector}) is satisfied.

\subsection{Examples}

The situation described in the two previous subsections can be realized 
for special kinds 
of bosonic matter models:
 
1.Scalar fields without potentials, including fields subjected to non 
linear constraints, like the $\sigma$ model.  The general coordinate 
invariant action for these cases has the form $S_{m} =\int L_{m}\Phi 
d^{D}x$ where $L_{m}=
\frac{1}{2}\sigma,_{\mu}\sigma,_{\nu}g^{\mu\nu}$.

2.Matter consisting of fundamental bosonic strings.  The condition 
(\ref{Sector}) can be verified by representing the string action in the 
$D$-dimensional form where $g_{\mu\nu}$ plays the role of a background 
metric.  For example, bosonic strings, according to our formulation, 
where 
the measure of integration in a $D$ dimensional space-time is chosen to 
be 
$\Phi d^{D}x$, will be governed by an action of the form:
\begin{equation}
S_{m} =\int 
L_{string}\Phi d^Dx,
L_{string}= -T\int d\sigma 
d\tau\frac{\delta^{D}(x-X(\sigma,\tau))}{\sqrt{-g}}
\sqrt{det(g_{\mu\nu}X^{\mu}_{,a}X^{\nu}_{,b})}
	\label{String}
\end{equation}
where $\int L_{string}\sqrt{-g}d^{D}x$ would be the action of a string 
embedded 
in a $D$-dimensional space-time in the standard theory; $a,b$ label 
coordinates in the string world sheet and $T$ is the string tension.  
Notice that under a scaling (\ref{Eg})  (which means that $g_{\mu\nu}
\rightarrow\lambda^{-1}g_{\mu\nu}$), 
$L_{string}\rightarrow\lambda^{\frac{D-2}{2}}L_{string}$ , therefore 
concluding that $L_{string}$ is a homogeneous function of $g^{\mu\nu}$ 
of 
degree one, that is eq.(\ref{Sector}) is satisfied, if $D=4$.

3.It is possible to formulate {\em the point particle model\/} of matter 
in a way 
such that eq.(\ref{Sector}) is satisfied.  This is because for the 
free falling point 
particle a variety of actions are possible (and are equivalent in the 
context of general relativity).  The usual actions are taken to be 
$S=-m\int F(y)ds$, where 
$y=g_{\alpha\beta}\frac{dX^{\alpha}}{ds}\frac{dX^\beta}{ds}$ and $s$ is 
determined to be an affine parameter except if $F=\sqrt{y}$, which is 
the 
case of reparametrization invariance.  In our model we must take 
$S_{m}=-m\int L_{part}\Phi d^{4}x$ with $L_{part}=-m\int 
ds\frac{\delta^{4}(x-X(s))}{\sqrt{-g}}F(y(X(s)))$ where $\int 
L_{part}\sqrt{-g}d^{4}x$ would be the action of a point particle in 4 
dimensions in the usual theory.  For the choice $F=y$, condition 
(\ref{Sector}) is satisfied.  Unlike the case of general relativity, 
different choices of $F$ lead to unequivalent theories.

Notice that in the case of point particles (taking $F=y$), a geodesic 
equation (and therefore the equivalence principle) is satisfied in 
terms of 
the metric $g^{eff}_{\alpha\beta}\equiv\chi g_{\alpha\beta}$ even if $\chi$ 
is 
not constant. It is interesting also that in the 4-dimensional case
 $g^{eff}_{\alpha\beta}$ is 
invariant under the Einstein symmetry described by eqs.(\ref{Eg}) and 
(\ref{Ephi}).  Furthermore, since for point particles the theory allows 
an 
Einstein sector, we seem to have a formulation where at macroscopic 
distances there will be no difference with the standard GR, but where a 
substantial difference could appear in micro-physics.  The theory could 
be 
useful in suggesting new effects that may be absent in GR and that 
could 
constitute non trivial tests of the NGVE-theory and of GR.  Deviations 
from GR may have also interesting cosmological consequences for the early 
Universe\cite{Ter}

In all the above cases 1,2 and 3, solutions where the field 
$\chi\equiv\frac{\Phi}{\sqrt{-g}}$ is constant, exist if $M=0$ and then 
solutions of the eq.(\ref{M-eq}) coincide with those of the Einstein 
theory 
with zero cosmological constant.  Theories of fermions (including 
fermionic 
string theories) appear also can serve as candidate matter models where 
there will be an Einstein-Cartan sector of solutions, as will be 
studied 
elsewhere.

\bigskip

\section{The Cosmological Constant Problem}

\subsection{The de Sitter solution in the context of the 
NGVE-theory}

If we allow non constant $\chi$, we expect to obtain {\em effects not 
present in 
Einstein gravity and cosmology}. For example, as we will see, in the 
NGVE-theory de Sitter space is realized in an unconventional way, where 
the de Sitter metric holds, but such de Sitter space is supported by the 
existence of a variable scalar field  $\chi$ which in practice destroys 
the maximal symmetry.

Effects of a non constant $\chi$ can be 
studied first in the case where there is no matter $(L_{m} = T_{\mu\nu} 
= 0)$.  If we require maximal symmetry of the space-time metric in such a 
case, 
the field $\chi$ must satisfy the condition $\chi_{,\mu;\nu} = c\chi 
g_{\mu\nu}$ where c is a constant.  Geometry allows\cite{Geroch} such 
$\chi$ only for the value of $c=-\frac{R}{12}$ and it is then possible 
to 
have maximally symmetric metric with any value of $R = -\kappa M$.  
For 
example a de Sitter solution of eqs.(\ref{Scal}) and (\ref{M-eq}) (taking 
$D=4$) 
is $ds^{2}=dt^{2}- a^{2}d\vec{x}^{2}$, where $a = a_{0}exp(\lambda t)$, 
$\chi = \chi_{0} exp(\lambda t)$ and $\lambda^{2}= \kappa M/12$.

Let us consider the point particle model ( considered at the end of section
3) which 
satisfies the Einstein condition. In this case the physics of the de 
Sitter space is described by geodesics with respect to the effective metric 
$g^{eff}_{\alpha\beta}$. Notice that such a metric corresponds to a power 
law inflation: $ds_{eff}^{2}= d\tau^{2}-\tau^{6}d\vec{x}^{2}$, where $\tau 
=\tau_{0}exp(\lambda t)$ and $\lambda^{2}= \kappa M/12$. By examining the 
physical metric $g^{eff}_{\alpha\beta}$ we notice that there are not the 
10 Killing vectors of the de Sitter space. We see that although the 
metric $g_{\mu\nu}$ is maximally symmetric, the physical geometry is not 
maximally symmetric.   

\subsection{Particle creation, possible instability of the de Sitter 
space and the Parker condition}

Some years ago, Parker\cite{Parker} suggested a possible mechanism, 
based on particle production in the early universe, for selecting a zero 
cosmological constant. The basic assumption Parker made was that of the 
existence of an underlying theory where the cosmological constant can be 
dynamically adjusted by a process based on something like the familiar 
`'Lenz's law``, which requires that the equilibrium condition be achieved 
after the cosmological constant relaxes to a certain value.

The idea is based on the fact that for a 4-dimensional homogeneous, 
isotropic cosmological background, 
there is not real particle creation for massless conformally coupled 
scalar fields (MCCSF) satisfying
\begin{equation}
	\Box\phi +\frac{R}{6}\phi =0
	\label{CSF}
\end{equation}

 If there are massless minimally coupled scalar fields (MMCSF) or 
gravitons, in general particle creation takes place. However, when 
considering particle creation effect in a given background metric with 
zero scalar curvature, the MCCSF and MMCSF theories behave in the same 
way\cite{Parker}. Also for such a background $R=0$, one  expects no graviton 
production\cite{Parker}. The existence of radiation with $T_{\mu}^{\mu}=0$ 
does not affect these conclusions\cite{Parker}.

\subsection{Realization of the Parker cosmological condition in the 
Einstein sector of the NGVE-theory}

In the context of the NGVE-theory, we have that states with $R=0$ exist 
among many other possible states, provided the integration constant $M$ 
is chosen zero and the matter lagrangian on shell equals zero (see 
eq.(\ref{IntM})). The last condition is satisfied, for example, for the 
case of massless fermions and for electromagnetic radiation with 
$T_{(em)\mu}^{\mu}=L_{em}=E^{2}-B^{2}=0$. Such a state may be realized 
apparently for the universe filled by ultrarelativistic matter. 

As Parker\cite{Parker} points out, a de Sitter space, although not 
satisfying the condition $R=0$, has a chance however of being stable 
due to its maximal symmetry. This seems in accordance with the 
calculations of Candelas and Raine\cite{Candelas}. In our case however 
even this possibility seems to be excluded. Indeed, for the de Sitter 
space in the NGVE-theory we have $\chi =\chi (t)\neq constant$ which 
means explicit breaking of maximal symmetry since for a maximally 
symmetric space a scalar field must obey\cite{Weinbook} the condition
$\partial_{\mu}\chi =0$. Therefore we expect that in the NGVE-theory 
the de Sitter space suffers from the above mentioned instability towards 
particle production. Only flat space-time
 allows the possibility 
of maximal symmetry since in this case $\chi$ may be constant, which 
is clear from eq.(\ref{Scal}).

It is interesting to observe that the Parker condition in the framework 
of the NGVE-theory, when applied to the above 
mentioned case of an early universe dominated by ultrarelativistic 
matter ( massless fermions and electromagnetic radiation with 
$T_{(em)\mu}^{\mu}=L_{em}=E^{2}-B^{2}=0$) is a particular realization of the 
Einstein condition(\ref{Sector}). Notice that once the Einstein condition 
is satisfied, the direct coupling of the $\chi$-field with matter 
disappears (see eq.(\ref{Scal})). When this decoupling does not exist, like 
for example is apparent from eq.(\ref{Noncons}), for the small matter 
perturbations around a de Sitter space, we expect a tendency for any 
homogeneous $\chi$ state to loose energy to inhomogeneous degrees of 
freedom (from the point of view of thermodynamics a transfer of energy from  
homogeneous degree of freedom of the $\chi$-field into inhomogeneous 
degrees of freedom is more preferable than the reverse since the 
inhomogeneous modes are more numerous). An effective 
way to describe this would be the introduction of a friction term in the 
equation of motion for $\chi$. We expect this would lead to the decay of 
the de Sitter space towards a Friedmann epoch with $M=0$ and $\chi 
=constant$. The nonequilibrium dynamics explaining the details of how the
relaxation of the cosmological constant is 
achieved, or in our case how the constant $M$ is changed to zero, is a 
subject for future studies.

\bigskip

\section{Discussion}

There are many directions in which research concerning the NGVE-theory 
could be expanded. A subject of particular interest consists of the study
of the NGVE-theory in 1+1 dimensions. In this case, the model gives 
equations that coincide with those of Jackiw and Teitelboim\cite{jack-tel} when
the constant of integration $\kappa M $ in the NGVE-theory is identified with 
the constant scalar curvature which is imposed on the vacuum solutions 
of that model \cite{jack-tel} and where our field $\chi$ plays the same 
role, in the equations, as the lagrange multiplier field in the 
Jackiw-Teitelboim model \cite{jack-tel}.

Another model that resembles the NGVE-theory  studied here is the non 
dynamical $\sqrt{-g}$ theory (NDSQR) ( for reviews, see articles of Weinberg
 and 
Ng\cite{reviews}). For the NDSQR-theory, also any de 
Sitter space is a solution of the theory and the constant curvature of 
the vacuum solutions also appears as an integration constant. The 
differences between the NDSQR-theory and the NGVE-theory are however very
obvious. To mention just some: (a) The NDSQR-theory does not exist as a 
non trivial theory in 1+1 dimensions, while the NGVE-theory does. (b) The 
4-D de Sitter solutions do not posses the maximal symmetry for the NGVE 
theory case (due to the fact that the scalar field $\chi$ is not 
constant) while maximal symmetry is respected in the NDSQR-theory where 
flat space and de Sitter space have the same symmetry. (c) In the 
NGVE-theory, the matter energy-momentum tensor is not covariantly 
conserved, while in the NDSQR-theory it is. (d) In addition, the 
NDSQR-theory is for all practical purposes indistinguishable from GR, 
while the NGVE-theory is really a new physical theory, which could 
contain an Einstein sector of solutions, but in addition there is the 
potential of finding out new gravitational effects.

We should also point out that a more thorough study of the infinite 
dimensional symmetry found here should be made. In particular the 
restrictions on the possible induced terms in the quantum effective 
action seem to be strong if the symmetries (\ref{infphi}) and (\ref{L})
remain unbroken after quantum corrections are also taken into account.
In particular, symmetry under the transformations (\ref{L}) seems to 
prevent the appearance of terms of the form $f(\chi)\Phi$ (except of 
$f(\chi)\propto\frac{1}{\chi}$) in 
the effective action which although is invariant under volume preserving 
transformations (\ref{infphi}), breaks symmetry  (\ref{L}). The case 
$f(\chi)\propto\frac{1}{\chi}$ is not forbidden by symmetry(\ref{L}) and 
appearance of such a term would mean inducing  a `'real`` cosmological
term, i.e. a term of the form 
$\sqrt{-g}\Lambda$ in the effective action. However, appearance of such a 
term seems to be ruled out because 
of having opposite parity properties to that of the action given in 
(\ref{Action}). Of course, in the absence of a consistently quantized theory,
such arguments are only preliminary. Nevertheless it is interesting to 
note that if all these symmetry arguments are indeed applicable, this 
would imply that the scalar fields $\varphi_{a}$ can appear in the 
effective action only in the integration measure, that is they preserve 
their geometrical role.  

The physical meaning of the symmetry (\ref{L}) is puzzling. To get a 
feeling of the meaning of this symmetry, it is interesting to notice that 
such transformation becomes non trivial if $L\equiv -\frac{1}{\kappa}R+L_{m}$ 
is {\em not\/} a constant, in contrast to what we have studied so far in 
this paper. For the derivation of eq.(\ref{IntM}), which implies 
$L=constant$, we have assumed that $\Phi\neq 0$. Allowing $\Phi =0$ in 
some regions of space-time should be equivalent to allowing for 
`'defects`` which could be string-like or take the shape of other 
extended objects. In such event, the symmetry (\ref{L}) would become non 
trivial 
precisely at the location of such defects, a situation that 
reminds us of the reparametrization invariance of theories of extended 
objects.

In this context, it becomes then natural to explore the possibility that 
all the matter could arise  as regions of space-time where $\Phi =0$, 
i.e. as defects described above. This could be a way to realize an idea 
of Einstein and Infeld\cite{Eininf} that matter should arise as singular
points from a 
pure gravitational theory. In their words\cite{Eininf}: `'All attempts to 
represent 
matter by an energy-momentum tensor are unsatisfactory and we wish to 
free our theory from any particular choice of such a tensor. Therefore we 
shall deal here only with gravitational equations in empty space, and 
matter will be represented by singularities of the gravitational field``.
In GR, however, such mechanism could lead only 
to singularities of the metric, i.e. black holes (if the cosmic 
censorship hypothesis\cite{Penr} is correct) . Although black 
holes 
are interesting objects, they are of course not  satisfactory candidates 
for the 
description of the matter we see around. Defects that appear in the 
NGVE-theory are singularities of the measure which are not necessarily 
singularities of the metric. Furthermore the existence of the symmetry 
(\ref{L}) suggests to us a close connection between these singularities 
and reparametrization invariance of extended objects as mentioned above.

Finally, the crucial question concerning the large distance behavior of 
the model should be analyzed in detail. In particular, the study of 
general conditions under which the theory contains an Einstein sector of 
solutions or a set of solutions which deviate very little from those of 
the Einstein theory must be studied. In this respect, let us recall that 
we have already reviewed a number of cases where the Einstein condition 
and Einstein symmetry hold. In these cases the theory is guaranteed to 
contain an Einstein sector of solutions.

However, it would be interesting to study situations where the answer is 
not so clear. We have here in mind cases where neither the Einstein 
condition or the Einstein symmetry are exactly satisfied, but that they 
could appear only in the long distance behaviour of the theory, without 
being satisfied by the underlying microscopic theory. This possibility is 
suggested by the fact that the point particle limit allows a formulation 
consistent with the Einstein symmetry. In the point particle limit 
however, the underlying microscopic (field theoretic) structure is 
`'integrated out`` and in this way disappears. This of course suggests 
that the integration of microscopic degrees of freedom could, under very 
general circumstances, lead to a macroscopic theory satisfying the 
Einstein symmetry. The answer to this last question will of course demand 
further research.

\bigskip   

\section{ Acknowledgements}

We would like to thank H.Aratyn, J.Bekenstein, K.Bronnikov, A.Feinstein, 
J.Friedman, A.Guth, A.Heckler, J.Ib$\acute{a}\tilde{n}$ez, C.Kiefer, R.Mann,
V.Mostepanenko, L.Parker, D.Pav$\acute{o}$n, J.Perez-Mercader, Y.Peleg and
H.Terazawa for interesting conversations.

\pagebreak

\end{document}